# Improvement of Brain MRI at 7T Using an Inductively Coupled RF Resonator Array

Akbar Alipour, Alan Seifert, Bradley Delman, Gregor Adriany *Member, IEEE*, and Priti Balchandani *Member, IEEE*.

*Abstract*— It is well known that magnetic resonance imaging (MRI) at 7 Tesla (7T) and higher magnets can provide much better signal sensitivity compared with lower field strengths. However, variety of commercially available ultra-high-field MRI coils are still limited, due to the technical challenges associated with wavelength effect, such as flip angle inhomogeneity and asymmetric transmit and receive RF field patterns. We aimed to develop a passive RF shimming technique using an inductively coupled RF resonator array to improve brain MRI at 7T, focusing of cerebellum. Method: an inductively coupled RF resonator array was designed for placing inside the commercial head coil to enhance the transmit field homogeneity and to improve the receive signal sensitivity. Each element of the array is a coupled-split-ring resonator (CSRR), which they are decoupled for each other using critical overlap technique. Electromagnetic (EM) simulations were used to optimize CSRR design parameters and determine the array configuration. The electrical characteristics of the CSRRs and decoupling level were evaluated using vector network analyzer. EM simulations and thermal test were also performed to evaluate RF safety. Phantom and ex-vivo MRI experiments were performed to evaluate the transmit efficiency and signal sensitivity in the presence of the array. The simulation and experimental data were compared for with/without the array to assess the array performance. Results: Simulations showed 2-fold improvement in transmit efficiency in peripheral region using the array. Simulated and calculated results determined SAR levels well below recommended limits. MRI experiments showed 2 to 4-fold improvement in Signal-to-noise ratio (SNR) and 2-fold improvement in contrast-to-noise ratio (CNR) in cerebellum. Conclusion: We modeled a wireless RF resonator array that can improve the transmit efficiency of the standard head coil and enhance the signal sensitivity at brain MRI without compromising RF safety.

## I. INTRODUCTION

Magnetic field strengths of the magnetic resonance imaging (MRI) systems are driven continuously beyond the clinical established field strengths of 1.5 and 3T by increasing the interest in neuroscience applications [1-5]. Ultra-high-filed (UHF) MRI generally refers to imaging at field strengths of 7 Tesla (7T) or more. In 2017, "Comformite Europeene" mark was given for a 7T MRI system indicated safety and environmental protection standards, later same year, food and drug administration (FDA) approved the first clearance for clinical 7T MRI [6-9]. Currently, imaging above 8T is available only on research protocols approved by an institutional review board and the informed consent of the subjects [10-11]. 7T and higher UHF magnets provide opportunities to satisfy the high demand of increased signal-to-noise ratio (SNR), detailed spatial information, and functional contrast [12-13]. With the available higher SNR at 7T, various studies performed high-resolution MRI imaging of the brain including skull base and cerebellum revealing cerebellar cortical layers. It is well known that MRI at 7T can provide much better signal sensitivity compared with lower field strengths. This can be used to reduce the scan time, while improving the spatial resolution required visualizing small sized deep features in the brain.

However, many UHF MRI experiments are designed for only describing individual structures of the brain in more detail without covering the whole-brain, specifically the central nervous system (e.g., cerebellum, spine, neocortex). Because the variety of commercially available 7T radiofrequency (RF) coils are still limited, due to the technical challenges associated with wavelength effect (high operating frequency), such as inhomogeneity of the transmitted magnetic field into the subject ($B_1^+$) and asymmetric transmit and receive RF field patterns for surface coils [14-20]. This is predominantly because of the undesired trade-off between high temporal and spatial resolution for large field-of-views (FOVs) and the difficulties introduced by UHF systems. The most commercial head coils designed for brain imaging at 7T

A. A, A. S, B. D and P. B are from Biomedical Engineering and Imaging Institute, Icahn School of Medicine at Mount Sinai, 1470 Madison Ave., 1St Floor, Room 119 New York, NY 10029 (e-mail: akbar.alipour@mssm.edu)
G.A is from Medical School, University of Minnesota, Center for Magnetic Resonance Research 2021 6th Street SE Minneapolis, MN 55455 (e-mail: adria001@umn.edu)

are used for imaging the specific region (i.e. skull base) with a steep gradient in signal often observed in the lower brain and other inferior areas of the head [21-22]. The most commonly commercially available 7T head coil is Nova 1Tx/32Rx coil (1 Transmit / 32 Receive, Nova Medical, Wilmington, MA) consists of a relatively short single channel transmit birdcage coil surrounding 32 channels receive array. This coil is designed principally for brain imaging with a limited FOV relative to commercial head coils used at lower field strengths, where the RF excitation is mostly transmitted with the system's large whole body birdcage coil [9]. In addition to the high RF field non-uniformity at UHF systems, the increased vulnerability of UHF imaging to susceptibility induced image distortions near air-filled cavities is also challenging, especially in cerebellum imaging. Its physical location in the posterior cranial fossa and anatomical diversity, combined with its small size, makes the cerebellum a challenging area of interest for UHF MRI. We will not address the susceptibility artifact in this study.

In particular, it is possible to handle the $B_1^+$ non-uniformity caused by the RF wavelength effect using active and passive RF shimming techniques. Parallel transmission (pTx) is an active RF shimming technique that significantly improved $B_1^+$ homogeneity in the human brain at UHF MRI systems [14-16]. To achieve whole-brain MRI, a generalized pTx design structure was introduced and signified its utility for covering whole-brain at 7T, specifically in the cerebellum and context acquisitions. The results indicated that pTx can significantly enhance $B_1^+$ uniformity across the entire brain compared with a single-transmit configuration (i.e., Nova 1T×32Rx coil). However, pTx systems reported high local specific-absorption-rate (SAR) compared with a single-transmit configuration [9].

The use of dielectric pads (DPs; high permittivity material with $\varepsilon_r$>50) as a passive RF shimming method in MRI has been proposed to reduce $B_1^+$ inhomogeneity, improve SNR, and increase transmit efficiency [23-30]. These significant benefits can be explained by the modified Ampere's law; attribution of displacement currents within the DPs that add to the local magnetic field. Recently, high permittivity ($\varepsilon_r$>100) composite materials constructed from calcium or barium titanate powders mixed with deuterium oxide for greater advantages in MRI applications [26]. A variety of MRI experiments have been performed using DPs in association with a commercial head coil at 7T. The DPs positioning on one or both sides of the head can improve SNR and signal homogeneity in the cerebrum. Although the using DPs in conjunction with the head coil displayed improved excitation, SNR, and coverage in the anterior portion of the cerebellum, it also resulted in a strong RF field gradient across the cerebellum in the anterior–posterior direction, consequently resulted in lower SNR and lower excitation in the posterior cerebellum when the DPs were applied [30]. Another limitation was the high thickness (≈ 2 cm) of DPs, which occupied considerable amount of the area inside the head coil.

Developing new RF pulses is another approach to overcome the $B_1^+$ inhomogeneity problem in UHF MRI. A matched-phase adiabatic RF pulse pairs was developed using the Shinnar Le-Roux algorithm in spin echo (SE) sequences to provide immunity to the inhomogeneous $B_1^+$ field at 7T [31]. The pulse pair was modified into a single self-refocused pulse to minimize the echo time. The self-refocused adiabatic pulses produced $B_1^+$ distribution that were substantially more uniform than those achieved using a standard SE sequence. This method is limited to SE sequences and may result in high-energy abortion rate. A method based on universal pulses was also presented to minimize the $B_1^+$ inhomogeneity in brain imaging at 7T [32]. This technique avoids systematic measurements of the RF and static field profiles for each subject, which are required in most pulse design protocols. Such pulses do not include the subject-specific field distributions but yet are aimed to significantly improve performance compared with the conventional RF shim modes. This method is specific for only Nova (8Tx/32Rx, Nova Medical) head coil and is limited for some sequences.

In this study we designed a wireless passive RF resonator array aimed to improve the transmit efficiency and signal sensitivity in a conventional head coil at 7T MRI. To realize this, the array was placed against the posterior and inferior portion of the head inside the head coil to improve whole-brain MRI focusing on cerebellum. We first investigate the design parameters and electrical characteristics of a single RF resonator, and then applied the optimized parameters to array modeling. We used electromagnetic (EM) simulations and experimental methods to investigate the technical potential of the array in improving transmit efficiency and signal sensitivity. MR experiments were performed using cadaver brain at 7T MRI system.

## II. MATERIAL AND METHODS

A single passive RF resonator was electromagnetically simulated and modeled. An array consisted of 10 RF resonators was constructed based on the optimized design parameters. The array was designed to provide: (a) improved transmit efficiency, and (b) enhanced receive signal sensitivity. Network Analyzer tests performed to measure electrical characteristics (Q-factor, resonance frequency, and decoupling values) of the device. SNR, $B_1^+$ map, and RF safety studies were performed using conventional head coil with and without the passive RF resonator array. Finally, the device imaging performance was evaluated using cadaver brain at 7T MRI.

*A. Theoretical Background*

Inductive coupling between the passive RF resonator tuned to the Larmor frequency and MRI coils results in local $B_1^+$ and MR signal enhancement [33-35]. This principle originates from; (a) inductive coupling between the resonator and RF excitation which leads to effective flip angle (*FA*) increasing (Fig. 1a) and (b) inductive coupling between the resonator and magnetization vector (*M*) during the reception which leads to receive-signal amplification (Fig. 1b). The transmit magnetic flux $\Phi_{rf}$ of the transmit coil inductively couples to the RF resonator, the reaction of the resonator with transmit flux leads to circulating current in the resonator, which results in a local flux $\Phi_{re}$. $\Phi_{re}$ is an additional flux provided by theresonator inductance that is added to the original transmit flux $\Phi_{rf}$. Therefore, the total flux $\Phi_t$ around the resonator can be written as:

$$\Phi_t = \left(\frac{R+iwL}{R+iwL+\frac{1}{iwC}}\right)\Phi_{rf} \quad (1)$$

at resonance, when $w = w_0 = \frac{1}{\sqrt{LC}}$,

$$\Phi_t = \left(1 + i\frac{wL}{R}\right)\Phi_{rf} = (1 + iQ)\Phi_{rf} = \Phi_{rf} + i\Phi_{re} \quad (2)$$

where $i$ represents a quadrature phase relationship between the transmit flux and the flux generated by the resonator. $Q$ is the resonator Q-factor.

The local flux generated by the resonator leads to an extra excitation field, which alters the effective *FA* by position. Based on the Lenz's law, the magnetic flux generated by the passive resonator is in opposite direction of the original flux, $\Phi_{rf}$, which may result in total flux, $\Phi_t$ cancelation at the very close vicinity of the resonator (e.g. at the center, where the flux generated by the resonator is strong). But at the outer side of the resonator $\Phi_t$ is amplified since the $\Phi_{re}$, and $\Phi_{rf}$ are in the same directions.

The fundamental signal in an MR experiment comes from the detection of the electromotive force (*emf*) for precessing magnetization [13]. The *emf* induced in the coil in our system can be expressed as:

$$emf = -\frac{d\Phi}{dt} = -\int B_t(\vec{r}) \cdot M(\vec{r},t)d^3r \quad (3)$$

where $B_t$ is the total magnetic field at a position (x,y,z) by unit current passing through the coil base on the principle of reciprocity [5]. $M$ is the magnetization vector.

According to the Faraday's law of induction, the nuclear magnetic resonance signal detected by the receiver coil in the presence of the RF resonator is:

$$S(t) \propto -\frac{d}{dt}\int B_t(\vec{r}) \cdot M(\vec{r},t)d^3r \quad (4)$$

from Eq. 2, $B_t = (1 + Q)B_{rf} \cong QB_{rf}$, therefore:

$$S(t) \propto -\frac{d}{dt}\int QB_{rf}(\vec{r}) \cdot M(\vec{r},t)d^3r \quad (5)$$

Additional local magnetic field generated by the resonator increases the signal intensity in the very close vicinity of the resonator by the factor of ($\approx Q$), which by moving away the increasing factor is decreased.

For an optimal signal enhancement the resonator normal axis should be in the same direction with the excitation filed. Tilting the resonator results in decreasing the coupling level, consequently signal amplification. In this study the resonators are almost in maximum coupling position.

### B. RF Resonator Modeling

A single RF resonator was simulated as a circular coupled-split-ring resonator (CSRR) [36]. The CSRR is a 3-layer structure; a flexible dielectric substrate ($\varepsilon_r$=3.4) which is sandwich between two metal Split-Ring Resonators (SRRs); the SRRs are 180°-rotated version (anti-oriented) of each other (Fig. 2a). SRR is an enclosed metal loop with a gap ($g$) along the loop [37].

When tuning a resonator, it is desirable to control the capacitance to reach the Larmor frequency, $f_{Lar}$ [36-38]. The built in distributed capacitance between the metal layers in CSRR structure used to tune the resonator to $f_{Lar}$ and avoids the need for any lumped element capacitance.

An equivalent circuit model of a CSRR is shown in figure 2b, where the resonator is modeled as a series RLC circuit with distributed capacitance and mutual coupling between the layers. Resonance frequency and Q-factor of a resonator are given by Eq. (1) and Eq. (2), respectively.

$$w_0 = \frac{1}{\sqrt{L_e C_e}} \quad (6)$$

$$Q = \frac{wL_e}{R_e} \quad (7)$$

where $L_e$ is the effective inductance, $R_e$ is an AC resistance of the structure, and $C_e$ is the effective capacitance of the overall structure. The approximate effective inductance, capacitance, and resistance of the given design can be formulated as [12]:

$$L_e = 2.54\mu D\left[\ln\left(\frac{2.07}{\rho}\right) + 0.18\rho + 0.13\rho^2\right] \quad (8)$$

$$C_e = \varepsilon_0 \varepsilon_r \frac{A}{d} \quad (9)$$

$$R_e = \frac{2l}{T\sigma\delta(1-e^{-b/\delta})} \quad (10)$$

where $\mu$ is the permeability of the copper, $D$ is the average diameter ($D = ((D_o + D_i)/2)$, $\rho$ is the fill ratio ($\rho = (D_o - D_i)/(D_o + D_i)$), $D_o$ is the outer diameter, $D_i$ is the inner diameter, $\varepsilon_0$ is the permittivity of the free space, $\varepsilon_r$ is the relative permittivity of the dielectric substrate, $A$ is the parallel plate surface area, $d$ is the distance between the consecutive layers (dielectric thickness), $W$ is the metallization (copper) width, $l$ is the path length of the metal trace, $b$ is the copper thickness ($35\mu m$), $\sigma$ is the conductivity of the copper, and $\delta$ is the skin-depth of the copper.

Electrical characteristics of the RF resonator rely on $L_e$, $R_e$, and $C_e$, which depend on four design parameters: (1) the average diameter $D$, (2) the dielectric thickness $d$, (3) gap $g$, and (4) the copper width $W$ (Fig. 1c).

A series of EM simulations (Computer Simulation Technology Microwave Studio (CST), Germany) was performed to investigate the effects of design parameters on the electrical properties of a single CSRR. Optimized resonator geometry was used for the array modeling.

## C. RF Array Modeling

A 10-element wireless passive RF resonator array was designed using 10 resonators (CSRRs). The elements were aligned in a form of a 2×5 matrix. The critical overlap technique was used to decouple adjacent elements [14]. We conducted EM numerical simulations (CST) to evaluate the EM field distribution of a head coil in the presence of the RF array. We modeled a head-sized quadrature-driven high-pass birdcage coil (22 cm in diameter and 25 cm in length), similar to the transmit head coil used in the experiments. The coil had 16 rungs connected at each end to two end rings and shielded by an open cylinder (23 cm in diameter and 27 cm in length). The coil was excited at two rungs, 90° apart in position and single phase, generating a circularly-polarized excitation. The coil was tuned by lumped capacitors distributed at the end-ring gaps and matched to 50 Ω using a single capacitor at each port, placed in series with an ideal voltage source with a 50 Ω internal resistance. For the simulation, the input power of the coil was adjusted to produce 1 W total power and a mean $B_1^+$ of 13 µT inside the head on an axial plane passing through the center of the coil. The performance of the 10-element RF resonator array was evaluated on a head model. The array was placed between the coil and the phantom: 3 cm away from the coil and 1 cm away from the phantom.

To calculate the maximum 10gr local average SAR values via simulation, we loaded the coil with a head model. The array was placed at the posterior position, between the head and the coil. RF excitation was performed using a birdcage head coil. Simulations were performed for both with and without the array. Numerical SAR results were acquired using a time domain solver and a power-loss monitor. Time-averaged SAR values were calculated by finding the time derivative of the incremental energy, absorbed by an incremental 10g mass of tissue. All resulting simulated SAR values were compared with the corresponding limits (10 W/kg for maximum local SAR and 3.2 W/kg for head average SAR) recommended by the FDA and IEC [39].

## D. Fabrication

A single RF resonator fabricated using the preferred design parameters found in the EM simulations. Parameter optimization was performed to obtain efficient electrical characteristics. The CSRR is a multilayer laminated structure consisting of two anti-oriented coupled SRRs, which are patterned on both sides of the dielectric substrate (Fig. 2a). The fabrication processes include the following steps: (1) a copper layer of SRR was patterned on one side of a flexible dielectric substrate (Kapton® polyimide films, DuPontTM). (2) a copper layer of SRR with 180° rotation was patterned on the other side of the substrate in the same axis with the first layer. The array was constructed using resonators with the following design parameters: $D_o = 50\ mm, D_i = 44\ mm, D = 47\ mm, \rho = 0.064, d = 100\ \mu m, T = 3\ mm, g = 18°, l = 152\ mm, A = 444\ mm^2$.

For RF array fabrication, 10 SRRs (first layer) were patterned on one side of a single piece of a dielectric substrate (Kapton), and then patterning 10 more SRRs (second layer) on the other side of the substrate completed the process. Second SRRs are in the same axis with first SRRs but with 180° rotation (anti-oriented). The total dimension of the array is 9 cm × 20 cm (Fig. 3a). The critical overlapping between neighbouring elements serves to reduce the inductive coupling between elements. The overlap is based on the critical loop center-to-center distance (0.76$D$) to minimize mutual inductance.

The built in distributed capacitance between two layers in a single CSRR was used for fine frequency tuning. Changing the conductor length can affect the capacitance and inductance values, consequently the operator frequency.

In order to prevent the over-flipping of the RF excitation and avoid boosting the absorption RF energy, some of the resonators were decoupled from RF excitation, specifically the resonators, which were in strong coupling position with RF excitation. The circuit model of a decoupled resonator is shown in figure 2a. The coupling level was evaluated using $B_1^+$ mapping method, which will be explained later. An antiparallel diode (Macom, Newport Beach, CA, USA) passively detuned the resonators during RF transmission.

## E. Electrical Bench Test

All elements were tuned to the Larmor frequency ($f_0$) at 7T (297 MHz) while loaded with the cylindrical saline phantom (15 cm in diameter and 30 cm in height; dielectric constant: 75; conductivity: 0.60 s/m). A foam pad (0.3 cm thick and $\varepsilon_0 = 2.1$) was placed between the array and the phantom to keep the same conditions as MRI experiments. Resonance frequency and Q-factor were assessed by measurements the S-parameters using a double pickup probe in a vector network analyzer (E5071C, Agilent Technologies, Santa Clara, CA, USA). Q-factor was calculated as a ration of a resonance center frequency to FWHM bandwidth in the transmission coefficient ($S_{21}$). Detuning performance of the resonators (detuned with antiparallel diode) was measured as the change in the $S_{21}$ of a double pickup probe. Decoupling between array elements was examined by $S_{21}$ measurements between the pairs of elements.

## F. Experimental Safety Analysis

To evaluate the effect of the RF resonator array in SAR distribution, temperature measurements were conducted in the vicinity of the array in an MRI scanner. The RF array was placed on top of a gel phantom (rectangle: 15 cm × 20 cm, $\delta_{gel}$=0.6 S/m, $\varepsilon_{gel}$=77), where a thin layer of plastic was used to avoid the direct contact of the array with the gel. The assembly was placed inside the head coil and was scanned for 15 min with a high SAR turbo-spin-echo (TSE) sequence (repetition-time (TR)= 1500 ms, echo-time (TE)= 8 ms, Flip-angle (FA)= 120°, bandwidth= 977 Hz/pixel, field-of-view (FOV)= 16×23 cm$^2$, matrix= 128×128). RF excitation was performed using a Nova1Tx/32Rx head coil (Nova Medical,

Wilmington, MA) in a 7T MRI scanner (Magnetom, Siemens Healthcare, Erlangen, Germany). Temperature was measured using four fiber-optic temperature probes (LumaSense Technologies, Santa Clara, CA) located at the high SAR value expected spots in the vicinity of the array. The temperature of a reference point far from the array was also collected. Baseline temperatures were recorded before RF transmission, and temperature changes were measured during scanning.

For the control experiment, which had no RF array, the local temperature rises at the temperature probe locations was determined. The probes were placed at the same spatial positions. SAR was calculated as:

$$SAR = C \frac{dT}{dt} \quad (11)$$

where $C$ is the heat capacity, $T$ is the temperature and $t$ is the time.

We visually examined the location of the probes relative to the RF array, immediately before and also after the heating assessment because significant variations in the measured temperature can occur due to slight variations in the probe positions relative to the array. Therefore, we used the exact same location of the probe when studying the temperature changes occurring with and without the array.

*G. Phantom MR Experiment*

The phantom (CuSo$_4$ solution) MR experiments were conducted to evaluate the array performance by characterizing the image SNR. The flexible and thin structure of the array allows to be placed on the curved surfaces to fully cover the interested region. All images were obtained on a 7T MR scanner using Nova1Tx/32Rx head coil. GRE (TR= 400 ms, TE= 4 ms, FA= 10°, bandwidth= 977 Hz/pixel, FOV= 16×21 cm$^2$, matrix= 128×128) sequences were used to compare the images acquired with and without the RF array. The combined system (commercial Nova 1Tx/32Rx head coil in combination with 10-element RF resonator array) performance was compared with a conventional Nova1Tx/32Rx head coil as a reference. SNR mapping was performed by obtaining two images with/without RF excitations, and then it was normalized by $B_1^+$ to isolate the receive sensitivity from the transmit field distribution.

We also evaluated the effect of the RF resonator array in transmit RF efficiency by mapping the $B_1^+$ field produced in the phantom using the double angle method. We used $B_1^+$ distribution to determined the coupling level between the coil and individual resonators.

The birdcage coil was not re-tuned and re-matched in the presence of the RF array, as tuning and matching were fixed for the commercial coil used in the experiments.

*H. Ex-vivo MR Imaging*

The ex-vivo MR imaging was performed in three cadaver brains (Musk Ox). The brains were fixed inside a cylindrical (12 cm in diameter and 17 cm in length) formalin-filled (400 mL of 10% neutral buffered formalin) container. The assembly was then vacuumed for 30 min to remove the bobbles.

The brains were imaged on a whole-body 7T MRI scanner (Magnetom, Siemens) using a single channel transmit and 32-channel receive (1Tx/32Rx) Nova head coil.

The RF array was placed at the posterior position of the head coil and the brain-contained container seated over the array. A foam pad with a thickness of 0.3 cm was used as an outer cover layer for the array.

MR images with and without RF array were obtained using GRE sequences. Following this, high-resolution T$_1$-weighted MP-RAGE sequences were applied. Proton density TSE sequences were also applied to assess the performance of the proposed RF array under various MRI sequences.

Ex-vivo contrast enhancement analysis was also conducted using calculation of contrast-to-noise ratio (CNR). CNR was calculated as:

$$CNR = \left| \frac{S_{ROI} - S_{REF}}{\sigma_N} \right| \quad (12)$$

where $S_{ROI}$ and $S_{REF}$ are the signal intensities of the region-of-interest (ROI) and reference point, respectively. $\sigma_N$ is the standard deviation of noise.

### III. RESULTS

*A. Resonator Modeling*

The inductive coupling between the transmit magnetic field and the RF resonator generates an additional magnetic field that manipulates the total magnetic field. In addition, in the receive phase, coupling between the magnetization vector (M) and the resonator enhances MR signal. The resonator inductance, $L_e$ plays a main role in coupling levels.

The overlapping area ($A$) and conductor length ($l$) are the major determinants in the $L_e$ and $C_e$ values of the resonator, which are controlled by the design parameters $D$, $W$, $g$, and $d$ values. The effect of design parameters was numerically analyzed for different $D$, $W$, $g$, and $d$ values. Results for five average diameter ($D$) values, with $d$, $g$, and $W$ kept constant ($d = 200 \mu m, g = 18°, W = 3\ mm$), showed that $L_e$ and $C_e$ were increased as $D$ increased (Fig. 4a). The average diameter increasing is associated with extending the $A$ and $l$, which results in increased $L_e$ and $C_e$ values, respectively.

The gap width ($g$) effect was studied with other parameters kept constant ($d = 200\mu m, D/W = 47/3\ mm$). As $g$ increased both $L_e$ and $C_e$ decreased, since the $A$ and $l$ decreased (Fig. 4b).

The dielectric thickness ($d$) effect was evaluated, with $D$, $g$, and $T$ kept constant ($D = 47\ mm, g = 18°, W = 3\ mm$). The effective capacitance $C_e$ decreased by increasing $d$, as the capacitance value is inversely proportional with the dielectric thickness (Fig. 4c). The effective inductance $L_e$ did not show a significant change by $d$ variations.

Results for conductor width ($W$) as the other parameters keep constant ($d = 200\mu m, D = 47, g = 18°$), showed that $C_e$ increased and $L_e$ decreased as $W$ increased (Fig. 4d).

Increasing $W$ results in higher overlapping area, consequently higher $C_e$ values. On the other hand, increasing $W$ leads larger conductor cross-section for electrical charge flow, which resulted in lower $L_e$ values.

Similar analyses were performed to evaluate the effect of the design parameters on resonance frequency ($f_0$) and Q-factor (Q). Increasing the $L_e$ and $C_e$ by increasing $D$ resulted in decreasing $f_0$ (Fig. 5a). Q decreased exponentially as D increased, which can be explained by the dominant effect of frequency decreasing. $f_0$ exponentially increased by increasing $D$ (Fig. 5b), science $L_e$ and $C_e$ were decreased. Q was not affected by $D$ variations. $f_0$ and Q linearly increased as d increased (Fig. 5c), this can be explained by decreasing $L_e$. $f_0$ exponentially decreased as $W$ increased (Fig. 5d), because of the dominant effect of decreased $L_e$. Q also decreased by increasing $W$. During each parameter evaluation, other parameters keep constant.

The following design parameters were used in 10-element array construction to have an efficient performance: $D_o = 50\ mm$, $D_i = 44\ mm, D = 47\ mm, \rho = 0.064, d = 200\ \mu m, T = 3\ mm, g = 18°, l = 132\ mm$. These parameters were selected based on: (a) optimized Q: to have sufficient signal enhancement and avoid signal saturation, (b) resonator size: keep the size big enough to avoid wavelength effect.

*B. RF Array EM Simulation*

We simulated a 10-element (2×5) RF array, where the elements were decoupled from each other using critical overlapping method. The mutual coupling between elements is minimized and the resonators are decoupled when $y = 0.76\ D$. Simulated scattering (S) parameters of the decoupled resonator pair show a transmission coefficient $S_{21}$ of -24 dB at 300 MHz.

In the presence of the 10-element RF array, simulated S parameters of the transmit coil (birdcage) show no significant changes in reflected and forward power. $S_{11}$ and $S_{22}$ remained bellow -22 dB and $S_{21}$ value showed an insignificant increase about 1% in transmitted power. The birdcage head coil was not re-matched and re-tuned due to these negligible changes in S parameters.

The simulation results showed improved $B_1^+$ efficiency at the regions covered by the array. $B_1^+$ was enhanced up 2-fold in the brain, as compared to the case without the array. The RF array resulted in 2-fold improvement in transmit efficiency in the cerebellum and 27% decrease in the coil center.

Maximum 10gr local and average SAR values were simulated using the human model from the library of the CST. SAR distribution was computed in the birdcage coil in the presence of the array. The results showed that the local SAR increased from 1.32 to 1.76 W/kg and average SAR increased from 0.88 to 1.12 W/kg.

FDA and IEC recommended corresponding limits for maximum local SAR is 10 W/kg and for head average SAR is 3.2 W/kg. All computed SAR values were well below theses limits.

*C. Bench-top Measurements*

Although the required design can be achieved through a simulation analysis, it can also be tuned on the bench-top based on $S_{11}$ and $S_{21}$ measurements. Specifically, to obtain the minimum decoupling condition, $y$ should be adjusted such that the frequency with minimum $S_{21}$ is equal to the $f_0$ .

An array of two decoupled resonators with the same size as the simulated is shown in figure 6a illustrating the bench-top experiments used for S-parameter measurements plotted in figure 6b. As expected, the critically overlapped ($y = 0.76\ D$) resonators are strongly decoupled from each other, with a transmission coefficient $S_{21}$ of −19 dB. The strong decoupling also makes the reflection coefficient ($S_{11}$) plot symmetrical around the $f_0 = 297\ MHz$.

A 10-element array consisted of 10 resonators (2 columns and 5 rows) was further built in which two kinds of coupling exist between the elements: (1) coupling between neighbouring elements of the same row and (2) coupling between neighbouring elements of different rows (Fig. 6c). The same overlapping distance as 2-element ($y = 0.76\ D$) is used to decouple the resonators from neighbouring elements. Decoupling between elements was examined in the presence of the phantom by $S_{21}$ measurements between pairs of elements while all other neighbouring elements were detuned using antiparallel diode.

In the array construction, resonator overlaps were adjusted to achieve acceptable decoupling level ($< -15$ dB) between elements (Fig. 6d).

The loaded Q-factor for the resonators was calculated as: $f_0/\Delta f$, where $\Delta f$ is the FWHM bandwidth of the measured $S_{21}$. The average loaded Q-factor of 21 was calculated.

We also tested the effect of bending on the 2-element array by 30°. Bending the array in the middle did not significantly change the S-parameters.

*D. Heating Experiment*

After 15 min of RF transmission, a maximum temperature increase of 0.7°C was experienced at the capacitive region (P$_2$) of the resonator relative to the counterpart point in control (without array) set up. This resonator was strongly coupled with the RF excitation (i.e. it was not detuned using antiparallel diode). Corresponding SAR gain of 1.2 was calculated at point P$_2$. Other points recorded almost identical temperature increases. Therefore, for safety scan, a maximum peak SAR reduced by factor of 1.2 is recommended.

The temperature increases were normalized relative to the reference point temperature. The temperatures at each position remained at raised levels for several minutes after RF transmission was turned off. This specifies that thermal convection in the gel was low, which suggests that the gel experiment overestimated in-vivo vascular conditions, where blood-flow leads to faster convective cooling.

*E. Phantom MRI*

Experimental calculations in phantom demonstrated an overall SNR distribution in the presence of the RF array. The array resulted in 3-fold SNR enhancement in the region-of-interest (ROI, outlined in white). SNR maps normalized by the excitation $FA$ maps resulted in a 76% enhancement in the receive-only SNR in ROI, suggesting that the improvement in the SNR is primarily due to the increase in the $FA$ with RF array.

The spatial distribution of the $FA$ maps ($B_1^+$ maps) in the experiments indicated a similar trend to simulations, with improved performance toward the inferior region when the RF array is present. The $FA$ averaged across all subjects showed a mean improvement of 1.6-fold in the peripheral region and a decrease of 0.3-fold at the center, while the average $FA$ in the phantom remained approximately the same.

*F. Ex-vivo Brain MRI*

Proofs of concept ex-vivo MRI experiments at 7T were conducted on 3 cadaver brains using a Nova 1Tx/32Rx head coil in conjunction with our proposed 10-element inductively coupled RF resonator array. Images obtained using GRE sequences indicate significant improvement in SNR at the brain, particularly in the skull base and cerebellum (Fig. 7a, b, numbers indicate the SNR values). High-resolution MP-RAGE images obtained in the presence of the RF array resulted in 2-fold SNR enhancement at thalamus (Fig. 7c, d). The proton density TSE images show up to 90% SNR improved at thalamus and peripheral regions in the presence of the RF array (Fig. 7e, f).

The analysis of the CNR in the ROI indicated that contrast is enhanced on all images in the presence of the array. As shown in MR images (Fig. 7), RF array resulted in 87% contrast enhancement on GRE images, 87% on MP-RAGE images, and 78% on proton density TSE images.

IV. DISCUSSION

We have designed and validated an effective approach to improve transmit efficiency and signal enhancement in 7T MRI systems. A wireless passive RF resonator array providing solution for $B_1^+$ inhomogeneity problem was constructed and tested. The array (9×20 $cm^2$) consisted of 10 elements, which they were aligned in a form of 2×5 matrix. Each element was a CSRR (circular, $D$ = 47 mm), where the elements were decoupled from each other using the critical overlap technique and were tuned to operate at the Larmor frequency of a 7T MRI scanner (297 MHz). CSRR design parameters were optimized using EM simulation (CST) and then applied for array modeling. To prevent $B_1^+$ over-flipping, strongly coupled elements were decoupled from RF excitation using anti parallel diode. A flexible thin film (200 μm) substrate was used in the array fabrication, which was sandwiched between two metal layers. Flexible architecture of the array increases its implementation in the various locations.

In this study we focused on using a wireless RF array in conjunction with a standard head coil to improve the whole-brain MRI at 7T by improving the receive signal sensitivity and transmit efficiency in the brain, particularly in the cerebellum. The transmit and receive inductive coupling of the RF array with the RF excitation and magnetization vector, respectively not only improve the $FA$ and receive sensitivity near the brainstem and cerebellum but also extend the anatomical coverage to visualize regions inherently far from the coil. In this work, we used a commercial head coil to demonstrate the improvement in coil sensitivity. The coil suffered from limited sensitivity at the temporal lobes and cerebellum, which was improved by placing the RF array near inferior regions of the head.

The array performance was evaluated using EM simulations, bench tests, and MRI experiments. We demonstrated, in both simulations and experiments, that the sensitivity and the transmit efficiency of a commercial head coil at 7T can be improved in the skull base and cerebellum using a passive RF array. This enhancement in SNR was used for improving whole brain imaging including the cerebellum, where the standard coil is limited due to poor transmit and receive sensitivity.

SAR distribution of the standard coil was manipulated in the presence of the RF array. The averaged and 10gr local SAR decreased with array in the center, but increased at the periphery of the coil, while producing greater transmit efficiency at the periphery and almost the same at the center. Transmit inductive coupling between the RF excitation and some of elements in the array could be considered as a major reason of SAR amplification. Temperature tests also reported a maximum local SAR gain of about 20% in the presence of the array. Although the simulated and measured local and average SAR values increased with addition of the RF array, they were below limits recommended by the FDA and IEC.

Our results showed that in addition to a higher SNR in the cerebellum in the presence of the array, we also observed higher SNR in the other regions of the brain. $FA$ maps and SNR calculations results were consistent with the obtained GRE, MP-RAGE, and TSE MR images, which showed improved visibility of the brainstem and cerebellum. These images also showed improved signal and contrast in the central and frontal regions of the brain. The experimental SNR analysis scaled by the $B_1^+$ determined that the achieved enhancement in SNR of 2- to 4-fold was mainly due to the improved transmit efficiency and partially (46%) due to received-only coupled sensitivity improvement. The CNR analysis of the images obtained with/without the proposed array also showed that the contrast was improved in the presence of the array.

V. CONCLUSION

In summary, our study demonstrates an efficient method to improve brain MRI focusing on cerebellum at 7T using a wireless passive RF array. The inductive coupling between the MRI coil and the array can improve the transmit efficiency and signal sensitivity as well as extending the anatomical coverage of the coil. The array consists of critically overlapped coupled-split-ring resonators. An EM simulation model was used to optimized resonators design parameters and electrical characteristics. The array safety assessments conducted in simulations and experiment showed that average and local SAR did not exceed current recommended limits. For this particular arrangement, the simulation and experimental results show an improvement in transmit efficiency, CNR and SNR, particularly in the cerebellum and inferior regions. This method could increase the feasibility of commercial head coils at 7T for whole brain MRI, functional MRI and other MRI applications. The future study of this work will focus of in-vivo validation in human. Although this device focuses on brain MRI at 7T, it can be modified to operate at different field strengths for various regions of interest imaging.

**Figures:**

Figure 1:

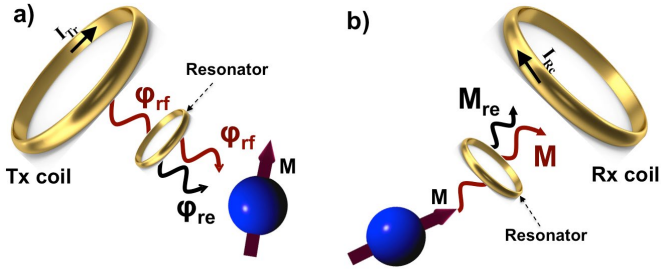

Fig. 1. (a) A passive RF resonator behavior during RF excitation, the resonator locally amplifies transmit field generated by transmit (Tx) coil and resulted in larger transvers magnetization. (b) During the receive phase, magnetization vector (M), which symbolized by a blue spin induces current ($I_{Rc}$) on receive (Rx) coil. Resonator coupling with M and Rx coil results in induced signal amplification.

Figure 2:

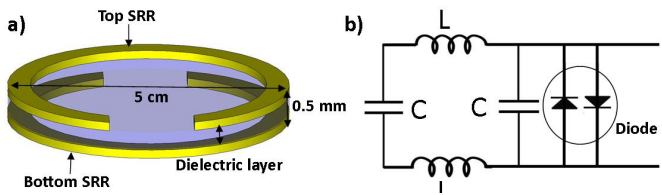

Fig. 2. (a) 3D schematic of a coupled-split-ring resonator (CSRR), which consists of two anti-oriented SRRs. A dielectric substrate is sandwiched between two layers of SRR. CSRR design parameters also are shown in 3D model, *D* is the average diameter, *W* is the metallization width, *d* is the dielectric thickness, and *g* is the gap width. (b) Electrical circuit model of the CSRR. C is the built in distributed capacitance value and L is the inductance. An anti-parallel cross diode is used to decouple the CSRR from RF excitation (diode is not shown in the 3D model).

Figure 3:

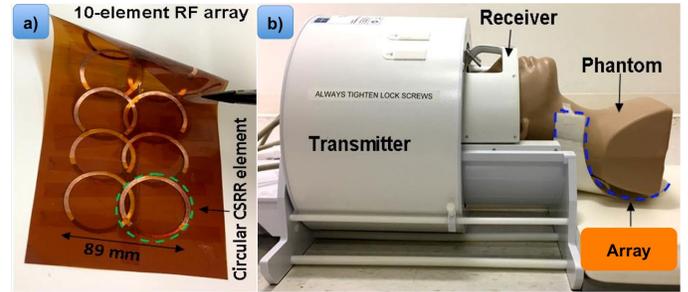

Fig. 3. (a) Prototype of the 10-element wireless passive RF resonator array constructed of 10 CSRRs. 10 SRRs are patterned on one side of the flexible substrate (Kapton), then after 10 anti-oriented version of these SRRs were patterned on the other side of the substrate to complete the CSRRs structure. The resonators were decoupled from each other using critical overlap technique. (b) The picture of a standard head coil used in our MRI experiments and position of the array relative to the coil and head phantom model.

Figure 4:

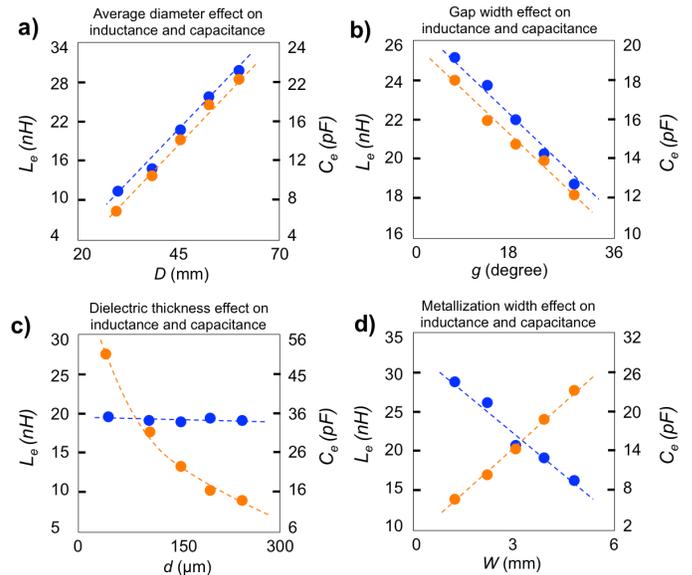

Fig. 4. EM simulation results detailing effects of the parameters *g*, *D*, *W*, and *d* on the resonator effective inductance ($L_e$) and capacitance ($C_e$). (a) Effect of gap width ($g$ = 6°, 12°, 18°, 24°, and 30°) with *D*, *W*, and *d* kept constant (D\W=47\3 mm, d=200 μm). The $L_e$ and $C_e$ decreased with *g*. (b) Effect of average diameter (*D*=27, 37, 47, 57, 67) with other parameters (*W* =3 mm, *d* = μm 200, and *g* = 18°) kept constant. As *D* increased, the $L_e$ and $C_e$ increased due to increasing the conductor length and overlap region. (c) Effect of metallization width (*W* = 1, 2, 3, 4, and 5 mm). The $L_e$ decreased because of larger conductor cross-section and $C_e$ increased due to increasing overlap area as the metallization width increases, with other parameters kept constant (*D* = 47 mm, *g* = 18°, *d* = 200 μm). (d) Effect of dielectric thickness (*d* = 50, 100, 150, 200, and 250 μm) with D\W=47\3 mm, *g* = 18°. $C_e$ decreased as it is proportional with the inverse of dielectric thickness. $L_e$ was not changed with *d* variations.

Figure 5:

Figure 6:

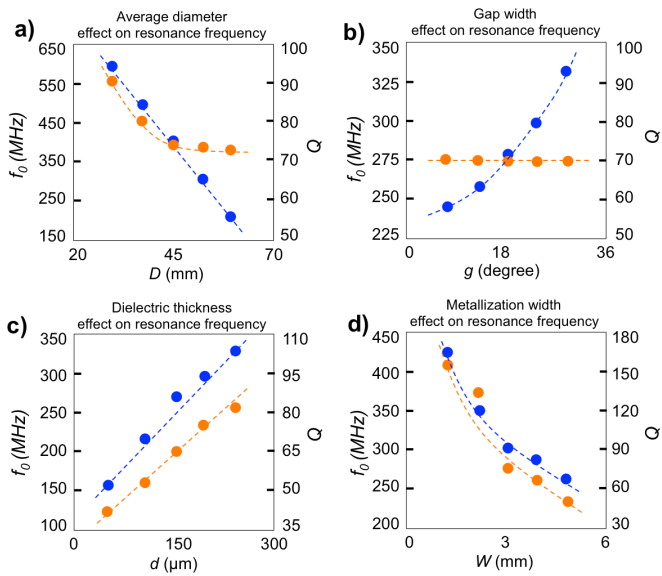

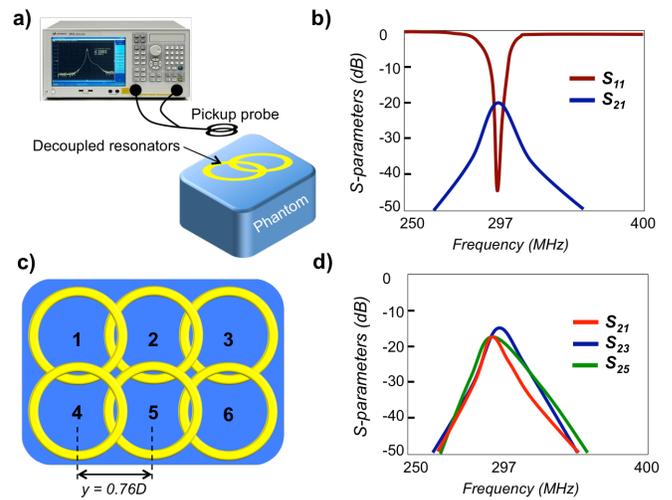

Fig. 5. EM simulation results detailing effects of the parameters *g*, *D*, *W*, and *d* on the resonator resonance frequency ($f_0$) and Q-factor. (a) $f_0$ decreased as *D* increased, which can be explained by the dominant effect of increased $L_e$. Q showed an exponential decreasing behavior as *D* increased. (b) $f_0$ exponentially increased and *Q* did not change as *g* increased. (c) $f_0$ and *Q* linearly increased as d increased. (d) $f_0$ and *Q* showed an exponential decreasing behavior as *W* increased. During each parameter evaluations other parameters were kept constant.

Fig. 6. (a) Bench test setup to measure the coupling between two resonators. (b) Measured S-parameters for two decoupled resonators (center-to-center distance = 0.76*D*, *D* is the average ring diameter). $S_{21}$ shows about -20dB decoupling level, the symmetric behavior of $S_{11}$ around $f_0$=297 MHz also proves the fact. (c) Schematic of the array to show the various coupling modes. All the elements were separated by critical overlapping technique (*y* = 0.76*D*). (d) Measured S-parameters show the coupling values of the resonator number 2 with two other resonators in the same raw (numbers 1 and 3) and one in the same column (number 5). Measured $S_{21}$, $S_{23}$, and $S_{25}$ show coupling values bellow -15dB.

Improvement of brain MRI using a passive RF array 11Figure 7:

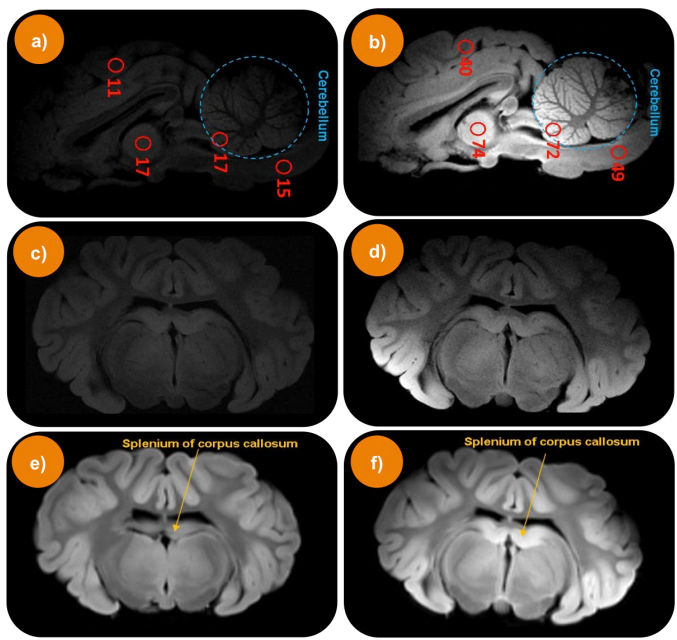

Fig. 7. Ex-vivo MR imaging using different sequences. First column (left) images (a, b, and c) were obtained without the array and second column (right) images (d, e, and f) were obtained with the array. Images were archived on 7T MRI scanner using a Nova 1-channel transmit and 32-channel receive (1Tx/32Rx) coil. Cadaver musk ox brains were used as the ex-vivo imaging model. Sagittal small tip angle GRE without array image (a) versus with array image (d) showing significant SNR and contrast enhancement in the whole – brain when using the array. In particular, skull base and cerebellum are more clearly visible in the presence of the array with about 4-fold SNR improvement. Numbers indicate the SNR values. Comparison of $T_1$-weighted coronal MPRAGE images (second raw, b and e) show signal and contrast improvement in the peripheral regions of the brain and thalamus when using the array. Proton density TSE images obtained without (c) and with (f) the array show signal enhancement of about 2-fold at thalamus.

**References:**

1. Uğurbil K. Imaging at ultrahigh magnetic fields: History, challenges, and solutions. *Neuroimage*. 2018;168:7-32. doi:10.1016/j.neuroimage.2017.07.007.

2. Uğurbil K, Adriany G, Andersen P, et al. Ultrahigh field magnetic resonance imaging and spectroscopy. *Magn Reson Imaging*. 2003;21(10):1263-1281. doi:10.1016/j.mri.2003.08.027.

3. Verma G, Balchandani P. Ultrahigh field MR Neuroimaging. *Top Magn Reson Imaging*. 2019;28(3):137-144. doi:10.1097/RMR.0000000000000210.

4. Grisoli M, Piperno A, Chiapparini L, Mariani R, Savoiardo M. MR imaging of cerebral cortical involvement in aceruloplasminemia. AJNR Am J Neuroradiol. 2005;26:657–661. PMID: 15760883.

5. Raichle, M. E. Images of the mind: Studies with modern imaging techniques. Annual Review of Psychology.1994 45(1), 333-356. doi: 10.1146/annurev.ps.45.020194.002001.

6. Zwaag W, Francis S, Head K, et al. fMRI at 1.5, 3 and 7 T: characterizing BOLD signal changes. NeuroImage 2009;47:1425–1434. doi.org/10.1016/j.neuroimage.2009.05.015.

7. Beisteiner R, Robinson S, Wurnig M, et al. Clinical fMRI: evidence for a 7T benefit over 3T. *Neuroimage*. 2011;57(3):1015-1021. doi:10.1016/j.neuroimage.2011.05.010.

8. Vu AT, Auerbach E, Lenglet C, et al. High resolution whole brain diffusion imaging at 7T for the Human Connectome Project. *Neuroimage*. 2015;122:318-331. doi:10.1016/j.neuroimage.2015.08.004.

9. Wu X, Schmitter S, Auerbach EJ, Uğurbil K, Van de Moortele PF. A generalized slab-wise framework for parallel transmit multiband RF pulse design. *Magn Reson Med*. 2016;75(4):1444-1456. doi:10.1002/mrm.25689.

10. Sadeghi-Tarakameh A, DelaBarre L, Lagore RL, et al. In vivo human head MRI at 10.5T: A radiofrequency safety study and preliminary imaging results. *Magn Reson Med*. 2020;84(1):484-496. doi:10.1002/mrm.28093.